\documentstyle[aps,epsfig,prd,preprint]{revtex}
\tightenlines

\def \beq{\begin{equation}}         \def \eeq{\end{equation}}
\def \beqa{\begin{eqnarray}}        \def \eeqa{\end{eqnarray}}
\def \bea{\begin{array}}        \def \eea{\end{array}}

\def \abs#1{\left| #1 \right|}
  
\def\plb#1#2#3{    { Phys. Lett. }{\bf B#1}, #3 (#2)}
\def\prd#1#2#3{    { Phys. Rev. }{\bf D#1}, #3 (#2)}

\def\prl#1#2#3{    { Phys. Rev. Lett. }{\bf #1} ,#3 (#2)}

\begin{document}
\title{Muon anomalous magnetic moment \\ in the standard model with two Higgs doublets }
\author{{Yue-Liang Wu and Yu-Feng Zhou}\\
  {\small Institute of Theoretical Physics, Chinese Academy of Sciences, \\
Beijing 100080,  China}}
\maketitle
\begin{center}
 \begin{minipage}{130mm}
\begin{center}{\bf Abstract}\end{center}

   { The muon anomalous magnetic moment is investigated in the standard model with
   two Higgs doublets (S2HDM) motivated from spontaneous CP violation. Thus
   all the effective Yukawa couplings become complex. As a consequence of the
   non-zero phase in the couplings, the one loop contribution from the neutral
   scalar bosons could be positive and negative relying on the CP phases.
   The interference between one and two loop diagrams can be constructive
   in a large parameter space of CP-phases.  This will result in a significant
   contribution to muon anomalous magnetic moment even in the flavor
   conserving process with a heavy neutral scalar boson ($m_h \sim$ 200 GeV) once
   the effective muon Yukawa coupling is large ( $|\xi_\mu|\sim 50$).  In general, the
   one loop contributions from lepton flavor changing scalar interactions become more important.
   In particular, when all contributions are positive in a reasonable parameter space of CP phases,
   the recently reported 2.6 sigma experiment vs. theory deviation can be easily explained even
   for a heavy scalar boson with
   a relative small Yukawa coupling in the S2HDM. }
\end{minipage}
\end{center}

\newpage


Recently, the E821 Collaboration at BNL has reported their
improved result on the measurement of muon anomalous magnetic
moment $a_\mu=(g_\mu-2)/2$ \cite{E821}.  The difference between
the measurement and the Standard Model (SM) prediction
\cite{SMg-2} is reported \
\beq
\Delta a_\mu=a^{exp}_\mu-a^{SM}_\mu
=426\pm165 \times 10^{-11},
 \eeq
 which shows a relative large deviation (2.6 $\sigma$) from the SM calculation. At $95\%$
confidence level, $\Delta a_\mu$ lie in the range
 \beqa
  113\times 10^{-11} \leq \Delta a_\mu \leq 749\times 10^{-11}.
 \eeqa
 As muon is about 210 times heavier than electron, it is expected that the
new physics effects on muon anomalous moment may be considerable.
A large amount of works have been made on checking the new physics
contributions to $\Delta a_\mu$ from various models. Among those,
the contributions from supersymmetric particles and scalar bosons
as well as extra dimensions seem to be the most attractive
ones\cite{models,Haber,Chang,Chou,tau}. In this paper, we shall
make a systematic analysis in the standard model with two Higgs
doublets (S2HDM) motivated from the study of origin and mechanism
of CP violation\cite{ww1,ww2,ylw}. In this kind of model, the
Higgs sector of SM is simply extended by including an additional
Higgs doublet with all the Yukawa couplings being real. After
spontaneous symmetry breaking, the origin of both fermion masses
and CP-violating phases can be attributed to the well known Higgs
mechanism with a single CP-phase between two vacuum expectation
values. Namely CP symmetry is broken spontaneously \cite{tdlee}.
Thus all the effective Yukawa couplings in the physical basis
become complex. The S2HDM has also been investigated from other
considerations\cite{2HDM1,2HDM2,2HDM3,2HDM4,2HDM5,2HDM6}. It will
be seen that though S2HDM is the simplest extension of SM, it can
provide a possible good explanation for the recently reported 2.6
sigma experiment vs. theory deviation of the muon anomalous
magnetic moment.

There are several versions of 2HDM. To avoid flavor changing
neutral current (FCNC) at tree level, some kind of discrete
symmetries are often imposed on the Higgs potential, this results
in two kinds of models which are usually called the 2HDM of type I
and I$\!$I \cite{model1-2}.  Note that by imposing the discrete
symmetry, the spontaneous CP violation is forbidden with two Higgs
doublets and all the Yukawa couplings have to be real.

Contributions to muon $g-2$ from the 2HDM of type II have been
discussed in Refs.\cite{Haber}.  It seems that the one loop
contribution is not large enough to explain the data. Even for a
large value of $\tan\beta\sim 50$, one still needs a very light
mass of the scalar $M_h\sim 5 $ GeV.  If both the scalar and
pseudo-scalar are included, due to the cancelation between them,
the situation will be even worse.  One then needs to consider two
loop contributions. It was found  \cite{Chang,Chou} that from the
two-loop diagrams of Barr-Zee type\cite{BZ}, the contributions
from pseudo-scalar is positive and could be larger than its
negative one loop contributions. Thus the net effects for
pseudo-scalar exchange become positive. Provided a sufficient
large value of the coupling $\tan\beta$, its contributions can
reach the experimental bound. By including the two loop diagrams,
the pseudo-scalar mass must be below $80$ GeV even when
$\tan\beta$ is large and around $50$ \cite{Chou}. To avoid the
cancelation between scalar and pseudo-scalar exchange, the mass
of the scalar boson has to be pushed to be very heavy ( typically
greater than 500 GeV).

An alternative way of suppressing the FCNC at tree level is to
take the smallness of the off-diagonal elements in the Yukawa
coupling matrices by considering approximate flavor or family
symmetries\CITE{2HDM2,2HDM5,ww1,ww2,ylw}. It has been
shown\cite{ww1,ww2,ylw} that through the spontaneous CP violation,
rich sources of CP violation including the KM-phase in SM  can be
induced from a single relative phase between the vacuum
expectation values of the two Higgs doublets. In the S2HDM, as
there are flavor changing scalar interactions for both neutral and
charged scalar bosons, the bounds on the neutral Higgs mass can be
released through the inclusion of the internal $\tau$ loop at one
loop level \cite{tau}.  Besides this, as it will be shown below
that the complex and flavor dependent Yukawa couplings in S2HDM
may completely change the interference between one and two loop
diagram contributions.  As a consequence, one and two loop
contributions can be both positive and provide significant
contributions to $\Delta a_\mu$.

In the S2HDM, after spontaneous symmetry breaking, the two Higgs
fields are given by \cite{ww1,ww2,ylw}
 \beqa
 \phi_1=\left(
\begin{array}{c}
  \phi^+_1 \\
{1 \over \sqrt{2}}(v\cos\beta e^{i\delta}+\phi^0_1)
\end{array}
         \right), \;\;
\phi_2=\left(  \begin{array}{c}
 \phi^+_2\\
  {1 \over \sqrt{2}}(v\sin\beta +\phi^0_2)
   \end{array}
         \right).
\eeqa where the phase $\delta$ is the relative phase between the
vacuum expectation values of the two Higgs doublets.  This phase
is the physical one as it can not be rotated away by a
redefinition of the  fermion phase. Thus in the physical basis
(i.e., mass eigenstate basis), all the effective Yukawa couplings
become complex, which leads to rich sources of CP violation. It is
more natural to use another basis for Higgs fields through the
recombination
 \beqa
 H_1 = \phi_1 \sin \beta e^{-i \delta}-\phi_2\cos\beta,  \mbox{$\;\;\;$} H_2 = \phi_2 \cos\beta
e^{-i \delta}+\phi_2\sin\beta
 \eeqa
  such that
  \beqa
  H_1=\left( \begin{array}{c}
   H^+ \\
    {1 \over \sqrt{2}}(R+iI)
    \end{array}
         \right), \mbox{$\;$}
H_2=\left(  \begin{array}{c}
 0 \\
  {1 \over \sqrt{2}}(v+H^0)
  \end{array}
         \right),
\eeqa
 where $H^+$ is the charged scalar  and $R$, $I$ and $H^0$
are the three neutral scalar bosons.  Here the neutral Higgs boson
$H^0$ plays the role as the one in SM with one Higgs doublet. The
additional two neutral scalar bosons $R$ and $I$ will lead to new
interactions beyond the SM.

 The Yukawa interactions between scalar bosons and leptons in the S2HDM can
be written as the sum of two parts ${\cal L}_1$ and ${\cal L}_2$,
where ${\cal L}_1$ contains only flavor conserving interactions and
${\cal L}_2$ contains flavor changing ones, i.e.
\beq
{\cal L}_Y=(\sqrt{2} G_F)^{1/2}({\cal L}_1+{\cal L}_2)
\eeq
with
\beqa
{\cal L}_1&=&\sqrt{2}
    H^+ \sum\limits^{3}_{i} \xi_{l_i} m_{l_i}  \bar{\nu}_L^i l_R^i
   +H^0 \sum\limits^{3}_{i} m_{l_i} \bar{l}_L^i l_R^i
   +(R+iI)\sum\limits_i^3 \xi_{l_i} m_{l_i} \bar{l_L^i} l_R^i +h.c,  \\
{\cal L}_2&=&(R+iI)\sum\limits_{i\not=j}^3 \mbox{\large
$\xi$}_{l_i l_j}  \bar{l_L^i} l_R^j+h.c, \eeqa In the above
expression, $\xi_{l_i}$ and $\xi_{l_i\, l_j}$ stand for the flavor
conserving and flavor changing Yukawa couplings which are in
general all complex numbers. For simplicity, we will neglect, in
the following considerations, the mixing among the three neutral
scalar bosons, but their masses are considered to be largely
split and the mass eigenstates are denoted by $h$ and $A$ respectively.

The phase effects of the Yukawa coupling has been discussed in the
literatures\cite{ww1,ww2,ylw,Keung,ylw2}.  The complex couplings
may lead to sizable CP violation in charged and neutral Higgs
boson exchanging processes in quark sector, such as $b\to
s\gamma$\cite{ww2,ylw2,bsg}, $K^0-\bar{K}^0$,  $B^0-\bar{B}^0$,
$B^0_s-\bar{B}^0_s$ and $D^0-\bar{D}^0$ mixings\cite{Wu-Zhou},
etc. In the lepton sector, it may result in large electric dipole
moment of leptons \cite{Keung,ylw,Iltan-EDM}.


%
In the case of effective real Yukawa couplings for neutral scalar
bosons, $h$ is purely a scalar while $A$ is a pseudo-scalar. The
contributions to muon anomalous moment from these two particles
always have different signs at both one loop and two loop level.
Namely, for scalar $h$-exchange, the contribution is positive from
one loop and negative from two loop. But for pseudo-scalar
$A$-exchange , the one loop contribution is always negative and
the two loop contribution is positive. As the experimental data
indicate that the deviation of $g-2$ from SM is positive.  The
large two loop contribution may lead to the conclusion that the
$A$-exchange must be dominant and the $h$-exchange must be made to
be negligible small.

However, in the general case, the situation can be quite
different. As all the couplings are complex in S2HDM, the Yukawa
couplings for neutral scalar bosons $h$ and $A$ contain both
scalar and pseudo-scalar type fermion interactions.  Further more,
the couplings are all flavor dependent, which is far away from the
2HDM of type I and I$\!$I, where the couplings are given by a
simple function of $\tan\beta$. The fact that one loop and two
loop contributions depend on different complex Yukawa couplings
provides an alternative possibility that they may not always
cancel each other. In a large parameter space, the two kind of
contributions may be constructive and result in a large value of
muon $g-2$

{\bf
The above discussion is also suitable for SUSY model \cite{Chang,Chou} as
the Higgs sector of it  is exactly like the 2HDM of type I$\!$I.  Recently, the
contribution from  two loop diagrams with charged Higgs in SUSY is
also discussed\cite{Geng}.  However, In the case of SUSY,  the main contribution may arise from 
the chargino-sneutrino and neutralino-slepton couplings at one loops\cite{susy-contr}.
}
 
\begin{figure}
\centerline{  \psfig{figure=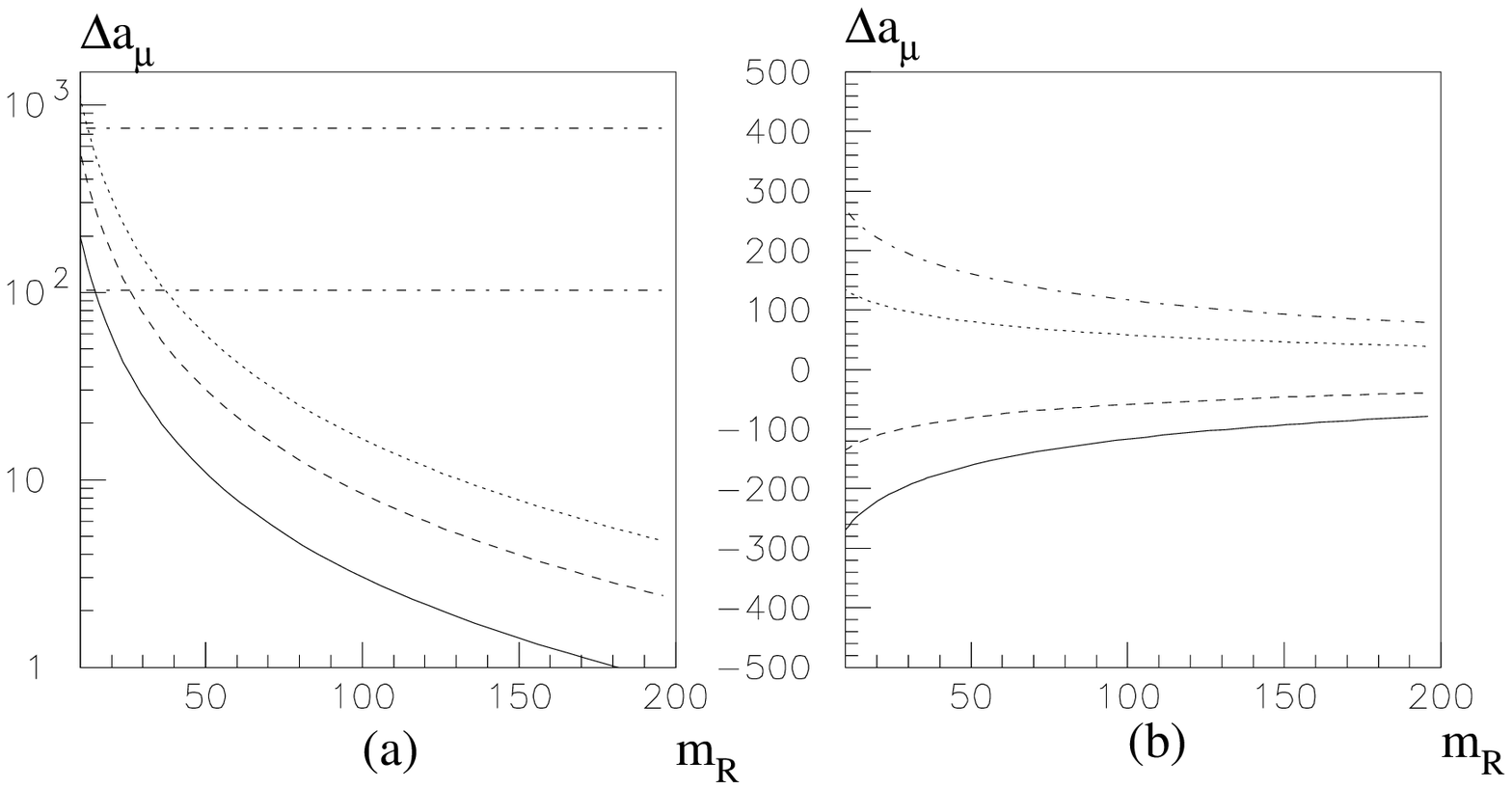, width=15cm}  }
\caption{\\
a) Single one loop flavor conserving contribution to $\Delta a_\mu$
as function of $m_h$. The solid, dashed and dotted cures correspond to
$|\xi_\mu|$=30,50,70 respectively.  The two dot-dashed horizontal lines
present the current experimental bound at $95\%$ confidence level \\
b)Single two loop contribution with $|\xi_\mu|$
being fixed at 50  and $\delta_\mu$ varies from 0, $\pi/3$, $2\pi/3$ to
$\pi$( from down to up). }
\label{oneloop.eps}
\end{figure}

 The one loop flavor conserving contributions from the scalar and pseudo-scalar
 have been investigated in Refs. \cite{oneloop}.
In S2HDM, the results are given by
\beqa
 \Delta a^h_\mu&=&{g^2\over 32 \pi^2} {m^4_\mu \over m^2_W m^2_h}
        \left[ (\mbox{Re}\xi_\mu)^2 \left( \ln {m^2_h\over m^2_\mu}-{7\over6}\right)
                      -(\mbox{Im} \xi_\mu)^2 \left( \ln {m^2_h\over m^2_\mu}-{11\over6}\right)
                \right]\\
\Delta a^A_\mu&=&{g^2 \over 32 \pi^2} {m^4_\mu  \over m^2_W m^2_A}
        \left[(\mbox{Im}\xi_\mu)^2 \left( \ln {m^2_A\over m^2_\mu}-{7\over6}\right)
                      -(\mbox{Re}\xi_\mu)^2 \left( \ln {m^2_A\over m^2_\mu}-{11\over6}\right)
                \right],
\eeqa
where $g$ is the weak gauge coupling constant. In the limit of
$m^2_\mu/m^2_h \ll 1$, the one loop results can be approximately
written as
\beqa
\Delta a^{h(A)}_\mu&=&\pm{g^2 \over 32 \pi^2} {m^4_\mu \over m^2_W m^2_h}
         \ln\left({m^2_h\over m^2_\mu}\right)\abs{\xi_\mu}^2 \cos 2 \delta_\mu
\eeqa
 where $\delta_\mu$ is the phase of $\xi_\mu$ with
$\xi_\mu=|\xi_\mu|e^{i\delta_\mu}$.  The one loop flavor
conserving contribution to $\Delta a_\mu$ is plotted in
Fig.\ref{oneloop.eps}a. It can be seen that its contributions
decrease rapidly with $m_h$ increasing, which means that the one
loop diagram cannot give large contributions to $\Delta a_\mu$
except for a very small value of $m_h\sim $  10 GeV or a very
large value of $|\xi_\mu|\sim $ 70. Note that in the S2HDM, the
contributions from $h$ and $A$  can be negative and positive
depending on the sign of $\cos 2\delta_\mu$, which is completely
different from the other type of 2HDM, such as type I and type
I$\!$I, where the $h(A)$ loop diagrams always give positive
(negative) contributions.

\begin{figure}
\vskip -8cm
\centerline{  \psfig{figure=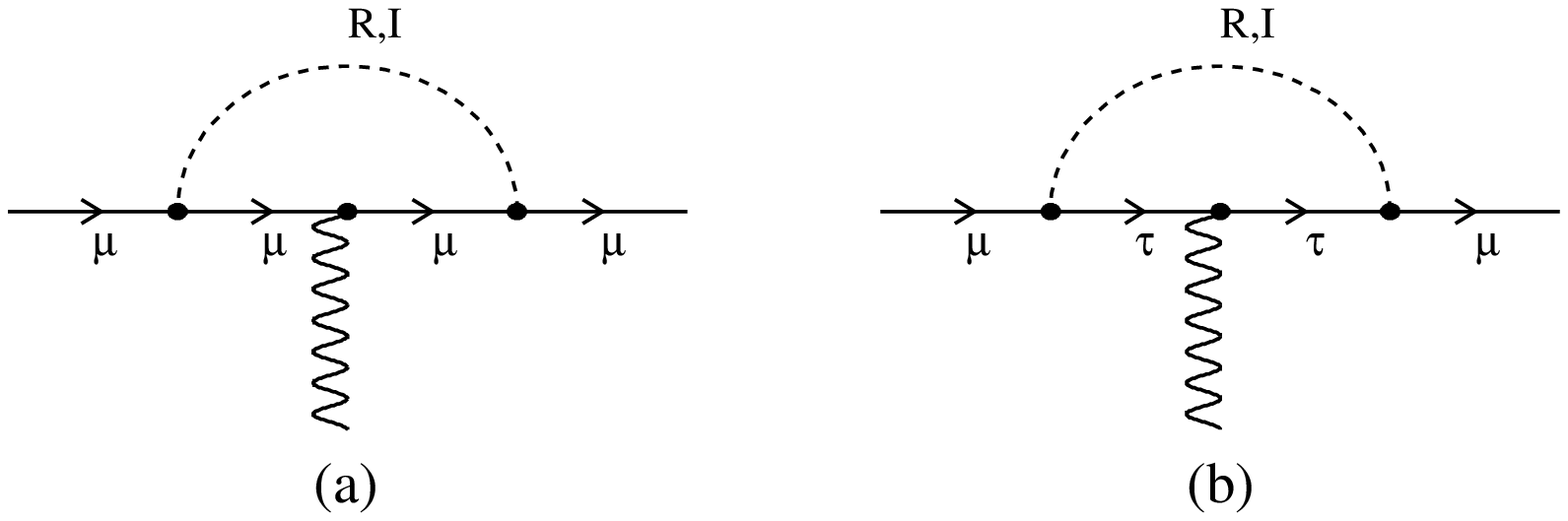, width=15cm} }
\vskip  -5cm
\centerline { \psfig{figure=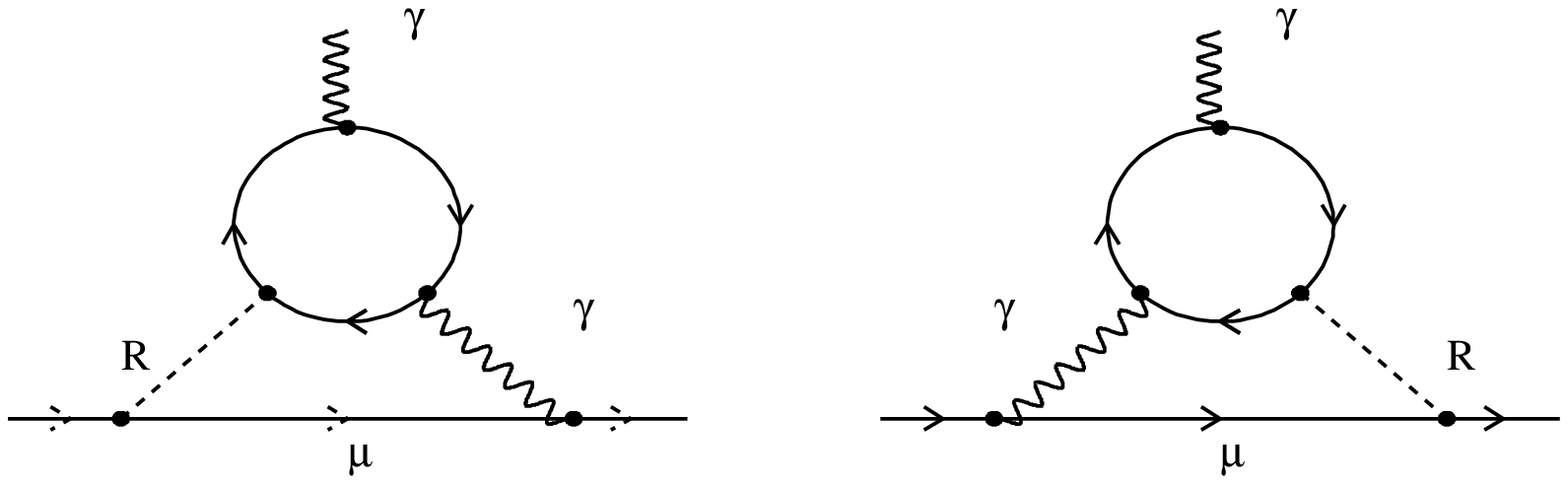, width=15cm}  }
\caption{Feynman diagrams contributing to $\Delta a_\mu$ in S2HDM.\\
(a) one loop flavor conserving.\\
(b) one loop flavor changing.\\
(c) and d) two loop diagrams of Barr-Zee type}
\label{1L.eps}
\end{figure}

In the case of one loop flavor changing processes ( see Fig.\ref{1L.eps}
b), the loops with internal $\tau$-lepton play an
important role as it is much heavier than the $\mu$-lepton.  This
then leads to an enhancement factor of the order
$m_\tau^2/m^2_\mu\sim {\cal O}(10^2)$ relative to the flavor
conserving one.  In S2HDM, the contributions from $h(A)$ exchange
are given by
\beqa
\Delta a^{h(A)}_\mu&=&\pm{g^2 \over 32 \pi^2}
{m^2_\mu m^2_\tau  \over m^2_W m^2_{h(A)}}
     \left(\ln {m^2_{h(A)}\over m^2_\tau}-{3\over2}\right)
     \abs{\xi_{\mu\tau}}^2 \cos 2\delta_{\mu\tau}
\eeqa
 In obtaining the above expression, we have taken
$\xi_{\mu\tau}=\xi_{\tau\mu}$ for simplicity.  The flavor changing
contributions to $\Delta a_\mu$ are presented in
Fig.\ref{oneloop.eps}b **with different values of $|\xi_{\mu\tau}|$.
The figure shows that the contributions to  $\Delta a_\mu$ may be
considerable large when $|\xi_{\mu\tau}|$ is large. For
$|\xi_{\mu\tau}|= 30\sim 50$,  the recently reported 2.6 sigma
experiment vs. theory deviation can be easily explained even for a
heavy scalar boson $m_h > 100$ GeV. So far, there are no strict
constraints on the values of the coupling $|\xi_{\mu\tau}|$.
Studies on the rare decays $\tau\to \mu(e) \gamma$ and  $\tau \to
3\mu(e)$ and the electric dipole moment of $\tau$ will be useful
to provide an interesting constraint on the parameter. However, as
the relevant experimental data at present are primitive and such
processes often contain more couplings such as $\xi_\tau$ and
$\xi_{\tau e }$ , the resulting constraints can not be clearly
obtained and they are not yet be very strong. To obtain the upper
bounds on $\xi_{\tau\mu}$, further studies are needed.

Now let us discuss the two loop effects on $\Delta a_\mu$. 
{\bf
Naively speaking, relative to one loop diagram, the two loop diagram
will receive an additional suppression  factor of ${ \alpha/4\pi \sim}{\cal O}(10^{-3})$ 
and is in general small.  However, there exists an
special kind of two loop diagrams which can give significant contributions
through the large Yukawa couplings between Higgs bosons and heavy
fermions. These processes which are often referred as Barr-Zee mechanism
 are shown in Fig.(\ref{1L.eps}.)$c$ and $d$
. In this kind of  diagrams, the Yukawa coupling will contribute a enhance factor of   ${m_f /m_\mu}$.
if $f$ is heavy quark such $t-$ quark,  it can reach the order of ${\cal O}(10^{3})$.
Thus the two loop contribution could be sizeable. 
The large Yukawa couplings of order  ${\cal O}(50)$  is need for explain the 
mass ratio between top and bottom quark in GUT\cite{GUT}.  Note that
arbitrary large Yukawa couplings may lead the theory to be non-perturbative .
This requires that the couplings have to be less than $1/(\sqrt{2} (\sqrt{2} G_f )^{1/2} m_f )$,
where $G_f$ is the Fermi constant.
For example,  for $t(b)$-quark, it is less than $1( 44)$. In the virtual  loop corrections,
as there are additional loop suppression factor of $(4\pi)^{-2}$. The constraints
may be weak. Further more, the loop integrals always decrease with Higgs mass
growing, and the Higgs loop contribution will decouple at large Higgs mass limit.   
}

\begin{figure}
\centerline{  \psfig{figure=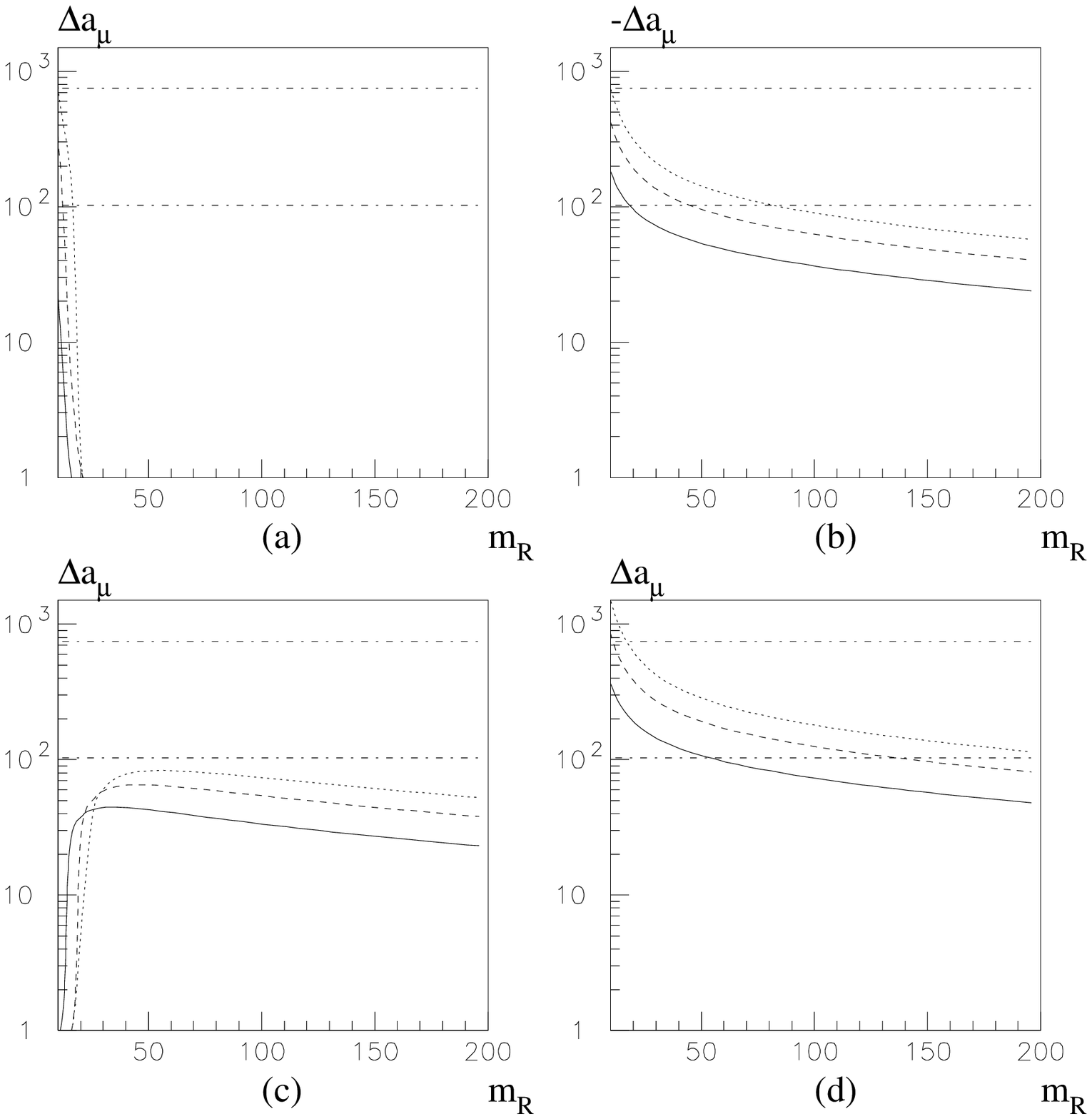, width=15cm}  }
\caption{The sum of the one loop flavor conserving and
the two loop contribution to $\Delta a_\mu$ as function of $m_h$.
for  $|\xi_t|=1$ and $\delta_t=0$.\\
(a) The solid, dashed and dotted cures correspond to $|\xi_\mu|$=30,50,70
respectively, with $\delta_\mu=0$.\\
(b) The same as (a) with  $\delta_\mu=\pi/3$. Note that all the contributios are negative.\\
(c) The same as (a) with  $\delta_\mu=2\pi/3$.\\
(d) The same as (a) with  $\delta_\mu=\pi$.
}
\label{twoloopa.eps}
\end{figure}

\begin{figure}
\centerline{  \psfig{figure=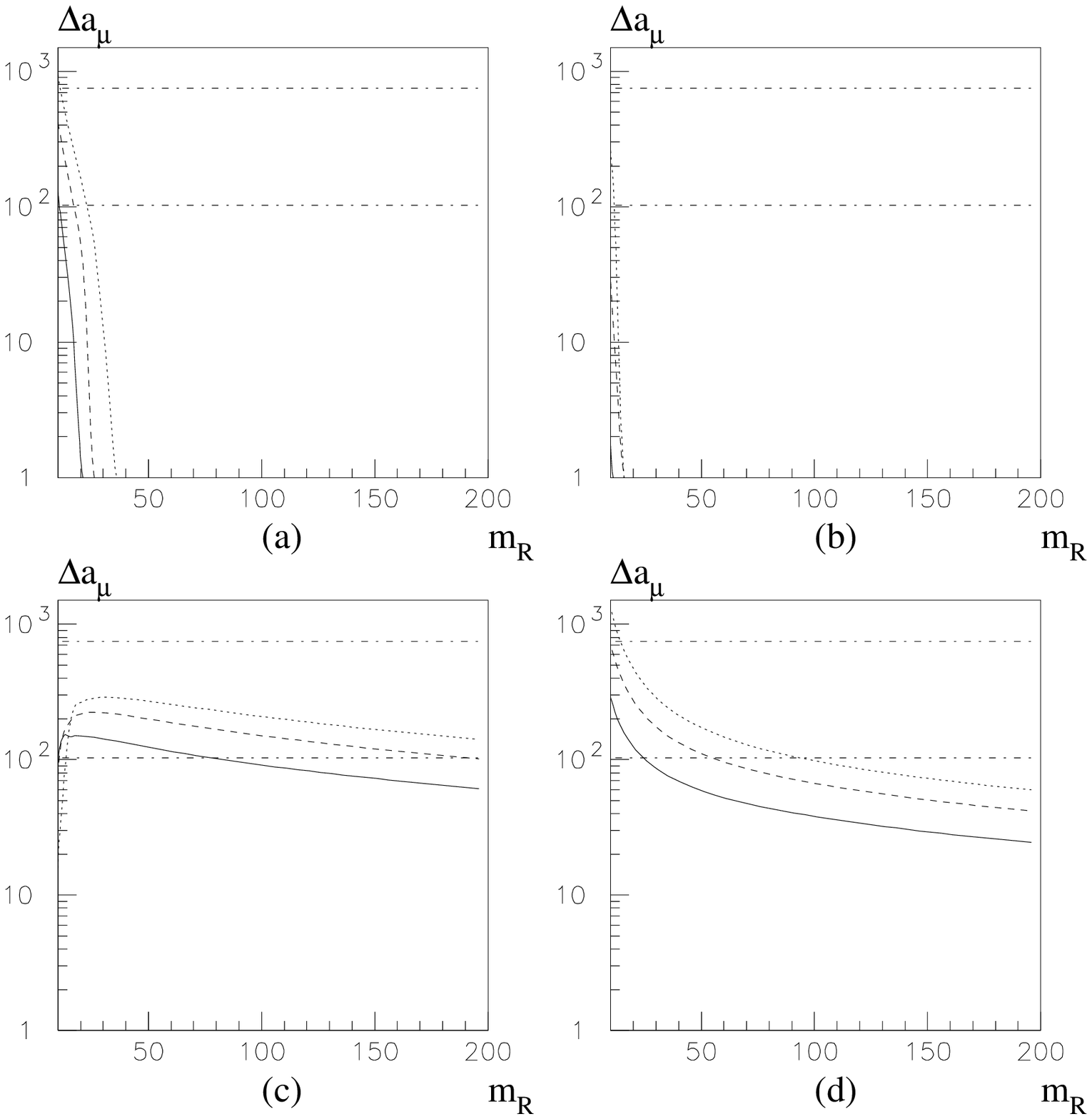, width=15cm}  }
\caption{The same as Fig.\ref{twoloopa.eps} for $\delta_t=\pi/3$}
\label{twoloopb.eps}
\end{figure}

For the case that the fermion
loop is dominated by top quark, the Barr-Zee type two loop diagrams
may lead to a sizable $\Delta a_\mu$. In S2HDM, their contributions
are given by \beqa \Delta a^h_\mu &=& { N_c q^2_t \alpha^2 \over
4\pi^2 \sin^2 \theta_W} {m^2_\mu \over m^2_W} \left[ \mbox{Im}\xi_t
\mbox{Im}\xi_\mu f\left( {m^2_t\over m^2_h }\right) -\mbox{Re}\xi_t
\mbox{Re}\xi_\mu g\left( {m^2_t\over m^2_h }\right) \right]\\ \Delta
a^A_\mu &=& { N_c q^2_t \alpha^2 \over 4\pi^2 \sin^2\theta_W} {m^2_\mu
\over m^2_W} \left[ \mbox{Re}\xi_t \mbox{Re}\xi_\mu f\left(
{m^2_t\over m^2_A }\right) -\mbox{Im}\xi_t \mbox{Im}\xi_\mu g\left(
{m^2_t\over m^2_A }\right) \right]
\eeqa
 where $N_c=3$ is the color number and
$q_t=2/3$ is the charge of top quark.  $\alpha=1/137$ and
$\sin^2\theta_W=0.23$ are the fine structure constant and weak
mixing angle respectively. $f(z)$ and $g(z)$ are two integral
functions which have  the following form \cite{BZ} \beqa
f(z)&=&{1\over2}z \int^1_0 dx {1-2x(1-x)\over x(1-x)-z} \ln
{x(1-x)\over z}\\ g(z)&=&{1\over2}z \int^1_0 dx {1\over x(1-x)-z}
\ln {x(1-x)\over z} \eeqa for large $z$, $f(z)\sim 1/3\ln z+13/8$,
$g(z)\sim 1/2\ln z +1$ and $f(1) \sim 1/2$, $g(1)\sim 1$.
The pure two loop contributions are plotted in
Fig.\ref{oneloop.eps}b. Unlike in the one loop case, where $\Delta
a_\mu$ decreases rapidly with growing $m_h$, the two loop
contributions decrease relative slowly and their signs depend on
the value of $\delta_{\mu\tau}$.  Therefore for a very large value
of $m_h$, the two loop effects become dominant. Another difference
is that in the one loop case the new physics contributions only
depend on $\xi_\mu$, while the two loop contributions depend on
two couplings $\xi_\mu$ and $\xi_t$, if the fermion loops for $b$
or $\tau$ are included \cite{Chang,Chou}, it will depend on more
parameters.  As the two couplings are complex numbers, the
interference between one and two loop diagrams may not always be
destructive. There exists a large parameter space in which the one
and two loop contributions are all positive. Thus they can result
in a large contribution to $\Delta a_\mu$ .  The constraint of
$|\xi_t|$ has been studied in $B^0-\bar{B}^0$ mixing and $b\to s
\gamma$ as well as the neutron electric dipole moment
\cite{Wu-Zhou,ww2,ylw,Keung,Wu-Zhou-beta}, the typical absolute
value for $|\xi_t|$ is of the order ${\cal O}(1)$. Taking
$|\xi_t|=1$ and $\delta_t=0$ as an example, the sum of one and two
loop contributions from $h-$scalar exchange reads
 \beqa
 \Delta a_\mu={\alpha \abs{\xi_\mu} \over \sin^2\theta_W}
                        { m^2_\mu \over m^2_W}
                        \left[  {\abs{\xi_\mu}\over 8\pi}
                                 { m^2_\mu \over m^2_h}\ln\left( {m^2_h\over m^2_\mu} \right)\cos 2\delta_\mu
                               - {\alpha\over 3 \pi^2} g\left({m^2_t\over m^2_h}\right) \cos\delta_\mu
                        \right].
\eeqa
 It is clear that if $\delta_\mu$ lies in the range $3\pi/4 \leq
 \delta_\mu \leq 5\pi/4$, the one and two loop contributions will be
 constructive.  The detailed numerical results for two different values
 of $\delta_t$ are plotted in Fig.\ref{twoloopa.eps} and
 Fig.\ref{twoloopb.eps}, where $\delta_\mu$ runs in a wide range.  It
 can be seen  that for large values of $\delta_\mu\sim \pi $, the
 contributions from $h-$scalar exchange can reach the experimental lower bound even for
 a heavy scalar with $m_h \sim 200$ GeV.  This is quite different from the
 existed 2HDM calculations in the literature, where the allowed range
 for the scalar boson $A$ must be less than 100 GeV , and the scalar boson
 $h$ must be much heavier than $A$, so that its negative
 contributions are negligible.

\begin{figure}
\centerline{  \psfig{figure=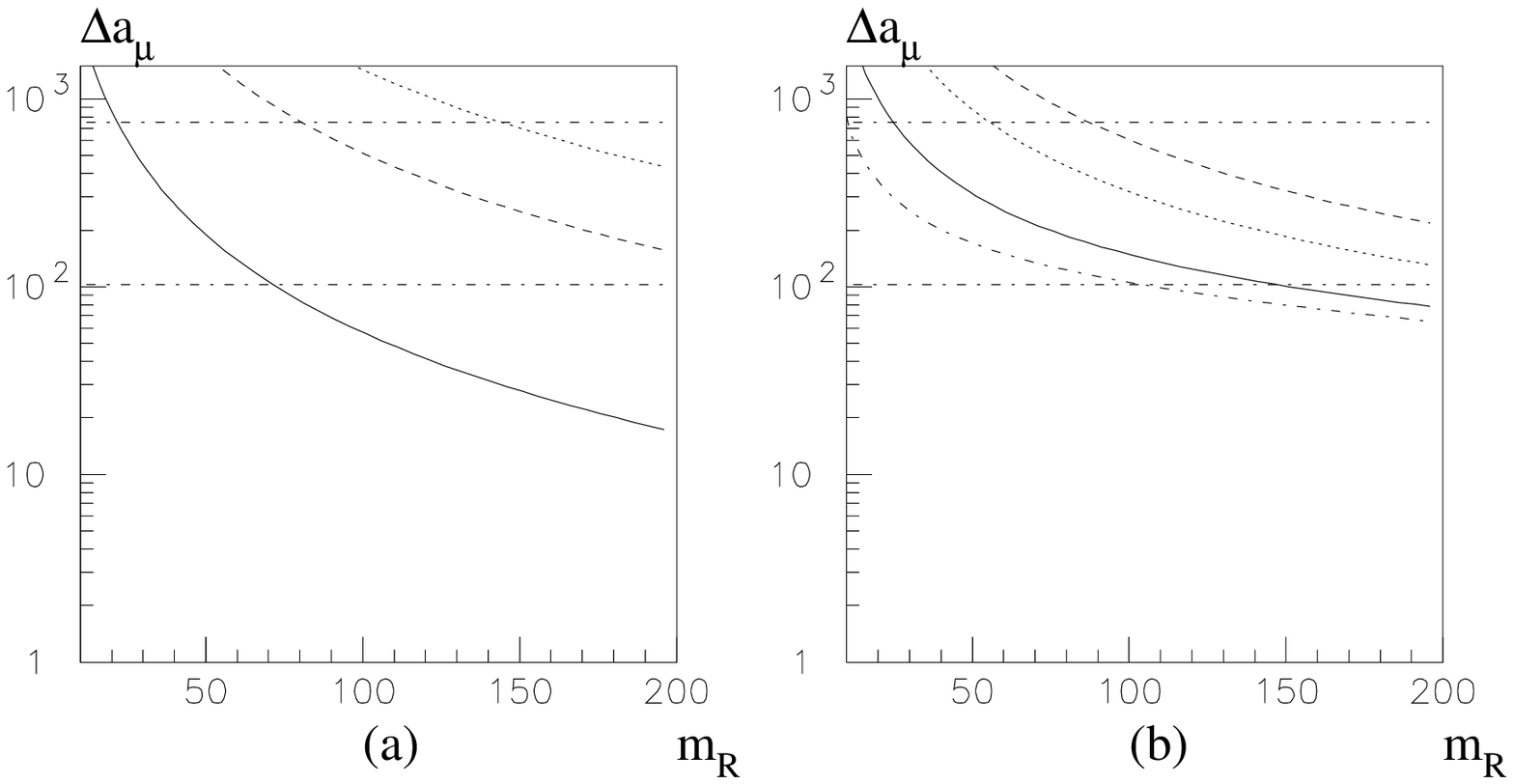, width=15cm}  }
\caption{\\
(a) The single flavor changing contributon, the solid, dashed
and dotted curves corresponds to $|\xi_{\mu\tau}|=$  10, 30,50,
with $\delta_{\mu\tau}=0$.\\
(b)The sum of one loop flavor conserving, one loop flavor changing and
the two loop contributions with $|\xi_t|=1$, $\delta_t=\pi/3$ and
$|\xi_\mu|=30$, $\delta_\mu=2\pi/3$.
The dot-dashed, solid, dotted and dashed
curves correspond to $|\xi_{\mu\tau}|=$ 5, 10, 20, 30,
with  $\delta_{\mu\tau}=0$.}
\label{tot.eps}
\end{figure}

In a large parameter space, it is also possible that contributions
from one loop flavor conserving and one loop flavor changing as
well as two loop diagrams are all positive. In this case, the
current data can be easily explained even with a relative small
value of $|\xi_{\mu\tau}|$.  The results for such a situation are
plotted in Fig.\ref{tot.eps}. It is seen that in this case a
reasonable value of $|\xi_{\mu\tau}|\sim 10 $ is large enough to
reach the experiment bound.

In conclusion, we have investigated all possible new contributions
to the muon anomalous magnetic moment from scalar interactions in
S2HDM. Though S2HDM is regarded as the simplest extension of SM,
it can provide a possible good explanation on the recently
reported 2.6 sigma experiment vs. theory deviation. In particular,
the important effects of the CP phases appearing in the Yukawa
couplings have been studied. Unlike the 2HDM of type I and I$\!$I,
the two loop $h-$ scalar contribution to $\Delta a_\mu$ could be
positive. The interference between one and two loop contributions
can be constructive and result in a considerable contribution to
$\Delta a_\mu$, so as to coincide with the recent E821 experiment
data.  In the case of constructive interference, the experimental
data $\Delta a_\mu$ can be understood even for a heavy scalar
boson $h$ with mass $m_h \sim 200$ GeV, once the Yukawa coupling
$|\xi_\mu|$ is large, $|\xi_\mu|\sim 50$. In general,
contributions from  one loop flavor changing diagrams are likely
to be the dominant one in a large parameter space. Especially, in
the case that contributions from all diagrams, namely, the one
loop flavor conserving and one loop flavor changing as well as two
loop diagrams in S2HDM with a reasonable parameter space, are all
positive, one can easily explain the current experimental data of
the excess of the muon anomalous magnetic moment.

\acknowledgments

This work was supported in part by the NSF of China under the
grant No. 19625514  as well as Chinese Academy of Sciences.




\end{document}